%% file: CSSforTelecom_5.tex
\documentclass[journal,12pt,onecolumn,draftclsnofoot,]{IEEEtran}
\usepackage{times}

\usepackage[noadjust]{cite}
\usepackage[dvips]{graphicx}
\usepackage{psfrag}
\usepackage{subfigure}
\usepackage{amsmath}
\usepackage{algorithmic,algorithm,indentfirst}
\usepackage{hyperref}
\usepackage{units}
\usepackage{color}
\usepackage{soul}
\usepackage[nolist]{acronym}
\usepackage{epsfig}
\usepackage{epstopdf}
\usepackage{caption}
\usepackage{tabularx}

\usepackage{xcolor}
\colorlet{activecolor}{blue}

\renewcommand{\baselinestretch}{1.87}
\begin{document}

\bstctlcite{IEEEexample:BSTcontrol}

\title{Complex Systems Science meets 5G and IoT}

\author{Nicola~Marchetti,
        Irene~Macaluso,
        Nicholas~Kaminski,
        Merim~Dzaferagic,
        M. Majid~Butt,
        Marco~Ruffini,
        Saul~Friedner,
        Julie~Bradford,
        Andrea~Zanella,
        Michele~Zorzi,
        and~Linda~Doyle
        \thanks{Nicola Marchetti (nicola.marchetti@tcd.ie), Irene Macaluso, Nicholas Kaminski, Merim Dzaferagic, M. Majid Butt,
          Marco Ruffini, and Linda Doyle are with CONNECT / The Centre for Future Networks and Communications,
          Trinity College, The University of Dublin, Ireland.
          Saul Friedner and Julie Bradford are with Real Wireless, UK.
          Andrea Zanella and Michele Zorzi are with the University of Padova, Italy.}%
        \thanks{This material is based upon works supported by the
          Science Foundation Ireland under Grants No. 13/RC/2077 and
          10/CE/i853.}  }

\maketitle
\vspace{-2.0cm}
\begin{abstract}
We propose a new paradigm for telecommunications, and develop a framework drawing on concepts from information (i.e., different metrics of complexity) and computational (i.e., agent based modeling) theory, adapted from complex system science. We proceed in a systematic fashion by dividing network complexity understanding and analysis into different layers. Modelling layer forms the foundation of the proposed framework, supporting analysis and tuning layers. The modelling layer aims at capturing the significant attributes of networks and the interactions that shape them, through the application of tools such as agent-based modelling and graph theoretical abstractions, to derive new metrics that holistically describe a network. The analysis phase completes the core functionality of the framework by linking our new metrics to the overall network performance. The tuning layer augments this core with algorithms that aim at automatically guiding networks toward desired conditions. In order to maximize the impact of our ideas, the proposed approach is rooted in relevant, near-future architectures and use cases in 5G networks, i.e., Internet of Things (IoT) and self-organizing cellular networks.
\end{abstract}
\vspace{-0.5cm}

\begin{IEEEkeywords}
\vspace{-0.5cm}
Complex systems science, Agent-based modelling, Self-organization, 5G, Internet of Things.
\end{IEEEkeywords}

\IEEEpeerreviewmaketitle

\input{acronyms}

\input{Introduction_5}
\input{Motivation_5}

\input{Methodology_5}
\input{CaseStudy_5}

\input{OpenChallenges_5}
\input{Conclusion_5}


\renewcommand{\baselinestretch}{1.35}
\bibliographystyle{IEEEtran}
\bibliography{CSSforTelecom}
\newpage

\end{document}

%% file: acronyms.tex
\begin{acronym}
%
%
%
%
%
\acro{3gpp}[3GPP]{3\textsuperscript{rd} Generation Partnership Program}
\acro{arq}[ARQ]{Automatic Repeat ReQuest}
\acro{awgn}[AWGN]{Additive White Gaussian Noise}
\acro{bcqi}[B-CQI]{Best \ac{cqi}}
\acro{bler}[BLER]{BLock Error Rate}
\acro{ca}[CA]{Carrier Aggregation}
\acro{comp}[CoMP]{Coordinated Multi-Point transmission and reception}
\acro{cqi}[CQI]{Channel Quality Indicator}
\acro{csi}[CSI]{Channel State Information}
\acro{dr}[DR]{Deployment Ratio}
\acro{dsl}[DSL]{Digital Subscriber Line}
\acro{edn}[EDN]{Extremely Dense Network}
\acro{eesm}[EESM]{Exponential Effective \ac{sinr} Mapping}
\acro{enb}[eNB]{evolved Node Base station} 
\acro{epc}[EPC]{Evolved Packet Core}
\acro{ffr}[FFR]{Frequency Fractional Reuse}
\acro{harq}[HARQ]{Hybrid Automatic Repeat reQuest}
\acro{icic}[ICIC]{Inter Cell Interference Coordination}
\acro{imt-a}[IMT-A]{International Mobile Telecommunications-Advanced}
\acro{irc}[IRC]{Interference Rejection Combining}
\acro{l2s}[L2S]{Link-to-System}
\acro{ll}[LL]{Link Level}
\acro{lte}[LTE]{Long Term Evolution}
\acro{lte-a}[LTE-A]{LTE-Advanced}
\acro{mac}[MAC]{Medium Access Control}
\acro{mcs}[MCS]{Modulation and Coding Scheme}
\acro{miesm}[MIESM]{Mutual Information Effective \ac{sinr} Mapping}
\acro{mimo}[MIMO]{Multiple Input Multiple Output}
\acro{ml}[ML]{Maximum Likelihood}
\acro{mmse}[MMSE]{Minimum Mean Squared Error}
\acro{mrc}[MRC]{Maximum Ratio Combining}
\acro{mu-mimo}[MU-MIMO]{Multi-User \ac{mimo}}
\acro{mui}[MUI]{Multi-User Interference}
\acro{nas}[NAS]{Non-Access Stratum}
\acro{nl}[NL]{Network Level}
\acro{oop}[OOP]{Object Oriented Programing}
\acro{pdcp}[PDCP]{Packet Data Convergence Protocol}
\acro{pf}[PF]{Proportional Fair}
\acro{phy}[PHY]{Physical layer} 
\acro{pmi}[PMI]{Precoding Matrix Indicator}
\acro{pon}[PON]{Passive Optical Network}
\acro{rb}[RB]{Resource Block}
\acro{rlc}[RLC]{Radio Link Control}
\acro{rr}[RR]{Round Robin}
\acro{rrc}[RRC]{Radio Resource Control}
\acro{rrh}[RRH]{Remote Radio Head}
\acro{rrm}[RRM]{Radio Resource Management}
\acro{sdr}[SDR]{Software Defined Networks}
\acro{snr}[SNR]{Signal to Noise Ratio}
\acro{sinr}[SINR]{Signal to Interference and Noise Ratio}
\acro{siso}[SISO]{Single Input Single Output}
\acro{sl}[SL]{System Level}
\acro{sic}[SIC]{Successive Interference Cancellation}
\acro{su-mimo}[SU-MIMO]{Single-User \ac{mimo}}
\acro{tbs}[TBS]{Transport Block Size}
\acro{ts}[TS]{Technical Specification}
\acro{tti}[TTI]{Transmission Time Interval}
\acro{ue}[UE]{User Equipment}
\acro{xgpon}[XG-PON]{10-Gigabit-capable Passive Optical Network}
\acro{zf}[ZF]{Zero-Forcing}

\end{acronym}

%% file: Introduction_5.tex
\section{Introduction}

The transition of humanity into the Information Age has precipitated the need for new paradigms to comprehend and overcome a new set of challenges. Specifically, the telecommunication networks that underpin modern societies represent some of the largest scale construction and deployment efforts ever attempted by humanity, with renovations occurring nearly continuously over the course of decades. This results in networks that consist of numerous subsections, each following its own trajectory of development, commingled into a complex\footnote{With the term 'complexity' we refer to a specific set of complex systems science quantities, related to the   interactions between network entities (rather than to entities themselves) and between networks. As the current and future trend is towards more diverse networks coexisting and more entities (e.g., within IoT, or ultra dense small cell networks), the amount of interactions will increase, leading to an increase in complexity (in the meaning given to the word by complex systems science).} cacophony.

A few emerging trends confirm the picture just drawn. Mobile and wireless networks are getting denser and more heterogeneous in nature. Nodes in the network vary hugely in form and functionality –- ranging from tiny simple sensors to sophisticated cognitive entities. There is a wider range of node and network-wide parameters to set, many of which are interdependent and which impact heavily on network performance. Networks are becoming more and more adaptive and dynamic, and many parameters are set during run-time in response to changing contexts. As networks evolve, all of the above issues become more exaggerated -– e.g., 5G networks will see more antennas, more base stations and devices, more modes of operation, more variability, and more dynamism. In a world like that, there is no way to systematically capture network behaviour. There is no straightforward network theory or information theoretic approach that can be used to describe the overall network or the interplay between the different networks.

We propose to tackle this by studying wireless networks from the perspective of Complex Systems Science (CSS), developing complexity metrics and relating them to more traditional measures of network performance. One of the key questions in CSS relates to the \emph{degree of organization} of a system \cite{Lloyd2001}, in terms of both the difficulty in describing its organizational structure (A), and the amount of information shared between the parts of the system as a result of the organizational structure (B). For example, the measure of excess entropy \cite{Feldman2003} (type (A)) can be used to describe the behaviour of a collection of self-organising networks \cite{Macaluso2014, Macaluso2016}; while the signalling complexity associated with future network resource management can be analyzed through a type (B) measure, i.e., functional complexity, introduced in \cite{Dzaferagic2017,Dzaferagic2017syscon}.

The above conceptual structure based on complexity informs an \emph{agent-based modelling (ABM)} paradigm to examine the interactions between the different entities that shape a network. ABM provides a method of modelling complex systems from the ground up, which allows for a deeper investigation of the interactions that shape the ultimate system performance. ABM provides powerful modelling of entities in a variety of areas and contexts \cite{Niazi2009, Cirillo2006, Tonmukayakul2005}. The attributes of ABM can be applied to inform communication networks' decision making; in particular, ABM can be used to investigate the impact of several Medium Access Control (MAC) component technologies on the Key Performance Indicators (KPI) of both telecom networks and applications, for example in the case of a wireless sensor network aiding an Internet of Things (IoT) system \cite{Kaminski2016}.

In summary, we propose a new paradigm for telecommunications, drawing on concepts of a complex systems science nature, to understand and model the behaviour of highly heterogeneous networks and systems of networks. We also employ our framework to create new technologies for supporting network operation. 

%% file: Motivation_5.tex
\section{Motivation}
\label{bsoa}

We propose the development of a conceptual framework as a means of exploring a broad range of possibilities in wireless networks, including a vast array of 5G technological possibilities. This framework for thought applies concepts from complex systems science \cite{Whitacre2010, Hooker2011} to provide a means to understand wireless networks holistically on a variety of scales. Specifically, we consider the communication patterns that enable network functions, by capturing all nodes necessary to perform a given function; then by drawing connections between these nodes we highlight their functional dependencies. We call a graph obtained in this way \emph{functional topology}. This approach allows us to analyze the communication patterns on multiple scales. The lowest scale models the communication between individual devices/nodes. In other words, the lowest scale focuses on the communication between a node and all the immediate neighbors of this node in the functional topology. The second scale models the communication between a node, all its immediate neighbors and all neighbors of its neighbors. The increasing scale size moves the focus away from the communication between individual nodes, and allows us to analyze communication patterns between groups of nodes (i.e., functional entities/groups).

Considering the high degree of heterogeneity and dense interplay of network elements in proposed 5G and IoT systems, achieving a holistic understanding of network operation is poised to become an even more challenging prospect in the near future. To address these challenges, we demonstrate the power of our framework for the modeling and analysis of relevant 5G scenarios, i.e., self-organizing cellular and IoT networks. While our framework supports innovation beyond these concepts, we feel these scenarios adequately represent the near-future applications of our work.

\begin{figure}
\centering
\captionsetup{justification=centering}
\includegraphics[width=0.8\linewidth]{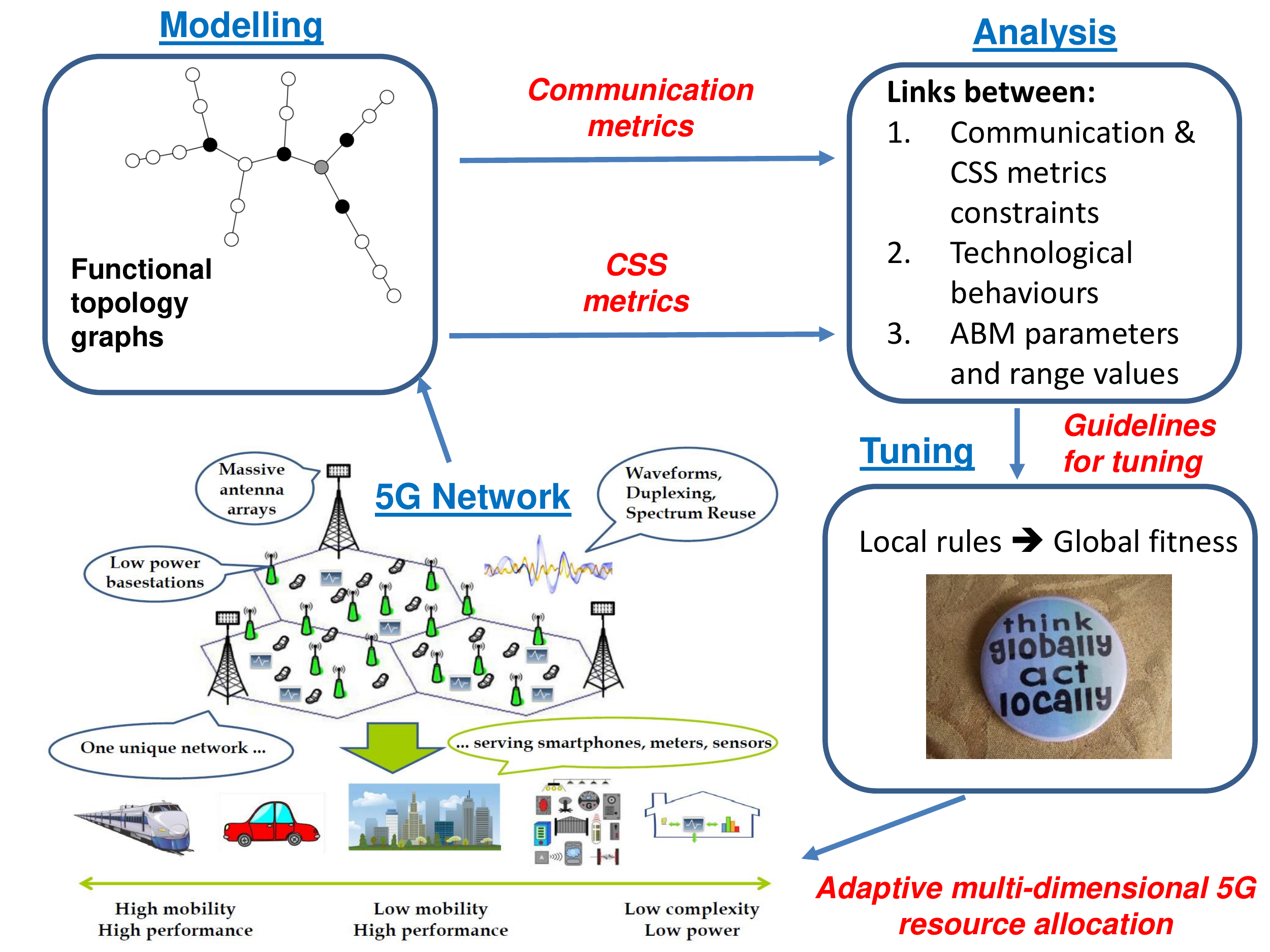}
\caption{\small{Our complex systems science based layered approach to 5G networks. Functional topology graphs are abstracted from the network, and are then used to compute complexity and telecom metrics, and find their relations. The understanding of such relations will then feed an ABM approach to network tuning.}}
\label{fig:concept}
\vspace{-0.75cm}
\end{figure}

The development of our concept is organized in a layered fashion, with a modelling layer forming the foundation of the framework and supporting analysis and tuning layers. The main aspects of our framework are represented in Fig. \ref{fig:concept} and will be discussed in detail in the remainder of the paper.

As compared to the CSS literature addressing communication systems \cite{Candia2008, Deville2014, Hidalgo2008, Onnela2007, Wang2009}, we study wireless networks from the infrastructure perspective. As a simple example, in \cite{Macaluso2014, Macaluso2016} excess entropy is used to measure complexity and, in combination with entropy, leads to an understanding of the structure emerging in a lattice of self-organising networks. The self-organising systems studied in \cite{Macaluso2014, Macaluso2016} exhibit a complex behaviour and this relates to robustness against changes in the environment; in particular, exploring frequency planning from a complex systems perspective leads to conclude that future networks shall eschew any current frequency planning approaches and instead determine frequency of operation on the fly. This has enormous implications for design and roll-out of networks, deployment of small cells, and network operation.

%% file: Methodology_5.tex
\section{Methodology}
\label{methodology}

Significant impacts have been made by CSS in a wide range of areas including physics, biology, economics, social sciences, computer sciences, and various engineering domains. We claim that the CSS perspective provides the necessary means to redefine the general understanding of telecommunication networks. We draw on concepts from information theory and ABM; each concept augmenting and developing the understanding of wireless networks. We will now briefly review some of the most important tools and concepts we use in our studies.

In order to specify and analyse the complexity of a network function, \cite{Dzaferagic2017} introduced a framework representing an abstraction of a telecommunication network, by modelling its operation and capturing all elements, i.e., nodes and connections, necessary to perform a given function. Our framework includes functional topologies, i.e., graphs created based on the functional connectivity between system entities (see Fig. \ref{fig:concept}). A node in our topology represents a functional entity of a network node or any information source that is part of the given network function. The links indicate dependencies between nodes. The definition of functional topologies allows us to visualise the relationships between system entities, and enables the systematic study of interactions between them. Based on these topologies one can define CSS inspired metrics such as functional complexity \cite{Dzaferagic2017}, which quantifies the variety of structural patterns and roles of nodes in the functional topology, or other information theoretical-inspired metrics.

Agent-based modelling (ABM) is a useful method to model networks. In \cite{Kaminski2016, Macaluso2014, Macaluso2016} ABM was used to investigate the impact of several MAC component technologies, in terms of both telecom and IoT application's Key Performance Indicators (KPI). This is key for our framework's analysis and tuning layers.

Our framework enables multi-scale modelling, analysis and tuning of wireless networks, in which changes in the 5G networks domain can be analysed and assessed. Indeed, in order to maximize the impact of our framework, our proposed approach is rooted in relevant, near-future architectures and use cases in 5G networks, such as self-organizing cellular and IoT networks. The use cases define the expected parameters, types of users/devices and environments; a general set of possible scenarios we could investigate using our framework is shown in Table \ref{table1}.

\renewcommand{\baselinestretch}{1}
\begin{table}[h]
\caption{\small{Possible use cases.}}
\vspace{-0.5cm}
\begin{center}
\begin{tabular}{||c | c ||}
\hline
\textbf{Parameters} & Low latency, High throughput, High reliability, Extensive coverage, \\
& Energy efficiency \\ [0.5ex]
\hline
\textbf{Type of users} & Typical mobile broadband, Healthcare, Automotive, Home/industrial automation, \\
&  Wearable devices   \\ [0.5ex]
\hline
\textbf{Environments} & Busy train station, Emergency/disaster location, Busy office complex/campus, \\
& Large utility/manufacturing plant  \\ [0.5ex]
\hline
\end{tabular}
\end{center}
\vspace{-0.85cm}
\label{table1}
\end{table}

\subsection{Solution Approach}


Our framework is based around the idea of using concepts, tools and measures of a complex systems science nature. The framework is based on a modelling layer which supports the analysis and tuning layers (see Fig. \ref{fig:concept}).

\subsubsection{Modeling Layer}

The modelling phase focuses on developing techniques to capture the significant attributes of networks and the interactions that shape them. Along with the traditional attributes used to characterize networks (e.g. coverage and throughput), the modelling phase develops new complexity metrics and investigates their relation to telecom KPIs. These metrics shall be developed distinctly for each application, based on existing and new concepts we draw from CSS.

The modelling component of the framework develops appropriate abstractions and formalisms to enable metric calculation. To this end, we produce a multi-scale abstraction for networks. The first level or device level of this abstraction focuses on individual elements within a network, targeting the interplay that results from information being collected and used locally by a single entity. Interference and stability of the connection (e.g., as a function of power available at the node) between nodes in a network are two examples of notions studied at the \emph{device scale}. Available local information may (as in the case of the interference perceived at a certain network node) or may not (battery level) result from the actions of other nodes. That is, the device scale typically models the implicit exchange of information, where nodes infer information for each other's actions without directly exchanging messages, such as the interaction-through-interference paradigm of a distributed Time Division Multiple Access (TDMA) system. The higher scales model the explicit exchange of information between groups of nodes in the network; at this level (\emph{interaction scale}) the nodes act on the basis of information provided by some other node directly, as occurs for example when assigning a slot in a centralized TDMA system.

\begin{figure}
\centering
\captionsetup{justification=centering}
\includegraphics[width=0.7\linewidth]{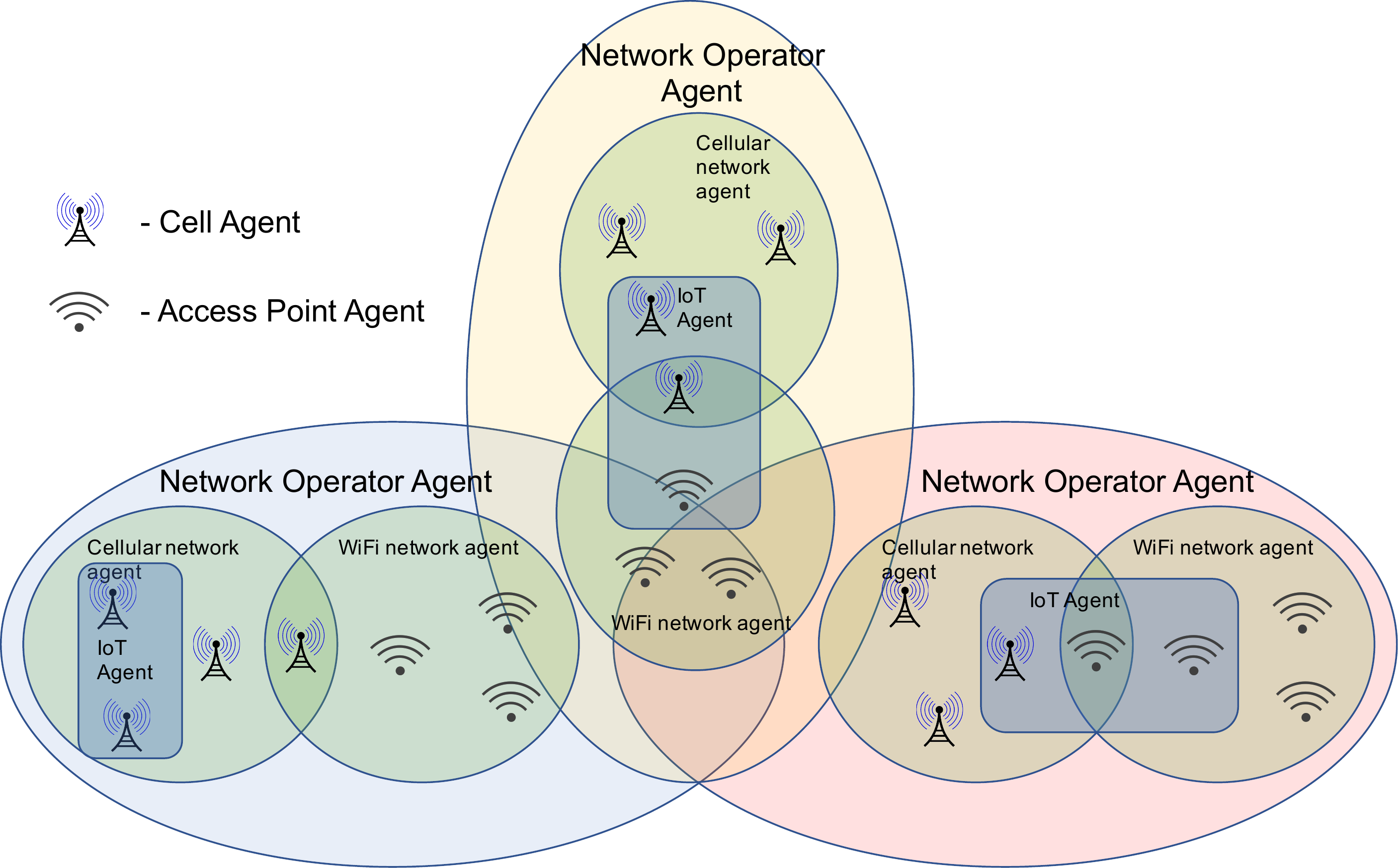}
\caption{\small{Agent organization. Our agent model is hierarchical, with major agents representing a whole network, and subagents representing IoT agents or individual cells and access points agents.}}
\vspace{-0.75cm}
\label{fig:agents}
\end{figure}

The interactions that shape the network formation and operation are directly modelled using ABM. Our model considers the interactions between the interests of different network operators. These agents operate in a hierarchical fashion (see Fig. \ref{fig:agents}) with network operator agents who, in turn, contain sub-agents that determine specific aspects of the network, based on technical behaviours. Anything that makes decisions in a network can be viewed as an agent, and ABM is applied to model interactions between agents. For example, IoT agents may attempt to use the infrastructure provided by operator agents, as shown in Fig. \ref{fig:agents}. To capture the range of possibilities, we can use nested subagents, in which major agents might represent a whole network with subagents representing individual cells. ABM allows conversion of experience with detailed processes (micro-level behaviours) into knowledge about complete systems (macro-level outcomes). In general we can consider several radio resources in our ABM model, e.g., resources belonging to frequency, power and space domains. Several alternative techniques and technologies can be applied within each domain, which entails a wide set of resources and related modes of utilisation.

\subsubsection{Analysis Layer}

In the analysis layer, the models are reviewed to determine the representative power and meaning of the metrics developed by linking: (i) the operator behaviours with our new CSS metrics; (ii) the operator behaviours with network KPIs (fitness); (iii) our new CSS metrics and KPIs. As an example, we could analyse the relationship between
operator decisions on the amount of shared resources (infrastructure and/or spectrum) and the resulting network characteristics.

For each scenario, measures of network performance can be identified, including standard operator KPIs such as cell edge, peak and mean throughput, spectrum utilisation relative to available bandwidth, network reliability, and coverage.

For each type of the above mentioned relations (i), (ii), (iii), we can determine the most promising pairing of elements (i.e., operator behaviour and CSS metric, or CSS metric and KPI) within each scale and between scales for determining connections. In particular, we can identify which behaviours correlate to specific network performance measures on each scale, and to what extent and how our CSS metrics describe these relationships. Further, we can investigate how a certain CSS metric-KPI relation at a certain scale affects another CSS metric-KPI relation at a different scale (e.g. a strategy leading to throughput maximisation at the device level might compromise the fairness objective of the resource allocation scheduler at the interaction level).

This process involves assessing the ability of the CSS metrics to describe the impact of operator behaviours, analysing the effect of these behaviours on the network KPIs, and finally describing the network KPIs in terms of the CSS metrics. Determining the link between network CSS metrics and KPIs would allow us to attempt to answer fundamental questions such as whether one needs a minimum complexity for achieving a given level of KPI (fitness), and what excess complexity implies in terms of adaptivity and robustness vs. cost.

In summary, the analysis layer completes the development of the core of our framework, by establishing a compact representation of the networks by linking complexity metrics to network performance.

\subsubsection{Tuning Layer}

The tuning layer augments the framework with algorithms that automatically guide the operation and management behaviours of relevant agents to achieve desired network properties. This tuning approach utilizes the holistic information encoded into the complexity based quantities, to select appropriate parameters and constraints for the behaviours of the agents. The developed tuning approach can be based on the application of multi-objective optimization techniques; the algorithms to be developed within this paradigm might apply multi-objective optimization algorithms (e.g., NSGA-II, PGEN, SMS-EMOA, successive Pareto optimization) to determine the Pareto fronts for the state spaces of the agent behaviours on the basis of achieving desirable CSS metrics values. These Pareto fronts provide the parameters and constraints of the operator behaviours, allowing operators to further optimize for specific differentiations while maintaining desired holistic properties. A particular solution may be selected from the Pareto front on the basis of agent preferences, such as a preference for high adaptivity and robustness or low complexity, without compromising the overall quality of the solution. 

%% file: CaseStudy_5.tex
\section{Applications of the Proposed Framework}
\label{applications}

\subsection{Modeling Layer}

\subsubsection{Agent-Based Modelling of the Internet of Things}

We employ an instance of our framework concept to investigate the tightened coupling between operative reality and information transfer precipitated by IoT. As such, this investigation resides primarily in the modelling phase with some extension into the analysis phase. Within this work, we apply the tool of ABM to study the impact of communications technologies within the scope of IoT \cite{Kaminski2016}.

\begin{figure}[h]
  \centering
  \includegraphics[width=8cm, height=5.5cm]{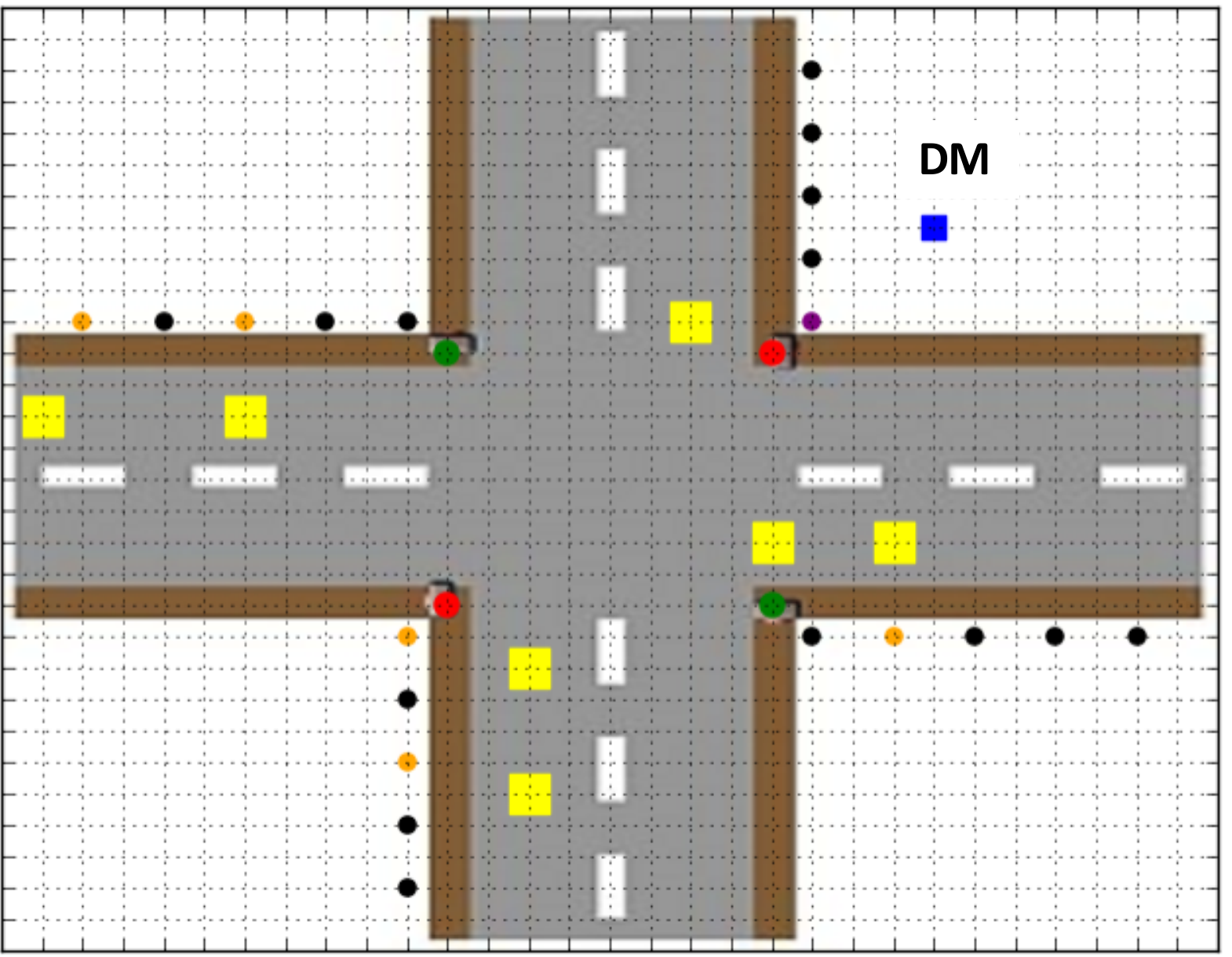}
  \caption{\small{Single Intersection Diagram. Sensors are deployed alongside the roads and are represented as dots. Inactive sensors are depicted as black dots; sensors detecting moving and static cars are shown as orange and purple dots respectively.}}
  \label{fig:intersection}
  \vspace{-0.5cm}
\end{figure}

An automatic traffic management system is considered, where for the purposes of illustrating the nature of our ABM approach, a single intersection is assumed, depicted in Fig. \ref{fig:intersection}, controlled with traffic lights, in which the avenue of the cross-road is observed by sensor nodes. A processing unit, here denoted as the decision maker (DM), serves as the sink of sensor information and the source of light control commands. Sensor nodes mark the advancement of cars, here portrayed as yellow squares and proceeding on the left side of the roadway, toward the intersection. Two MAC protocols (CSMA and Aloha) are investigated, for communication between the sensors and the DM. The DM applies the resultant information from this process to govern vehicular progress through the coloration of traffic signals.

\begin{figure}[h]
  \centering
  \includegraphics[width=9cm]{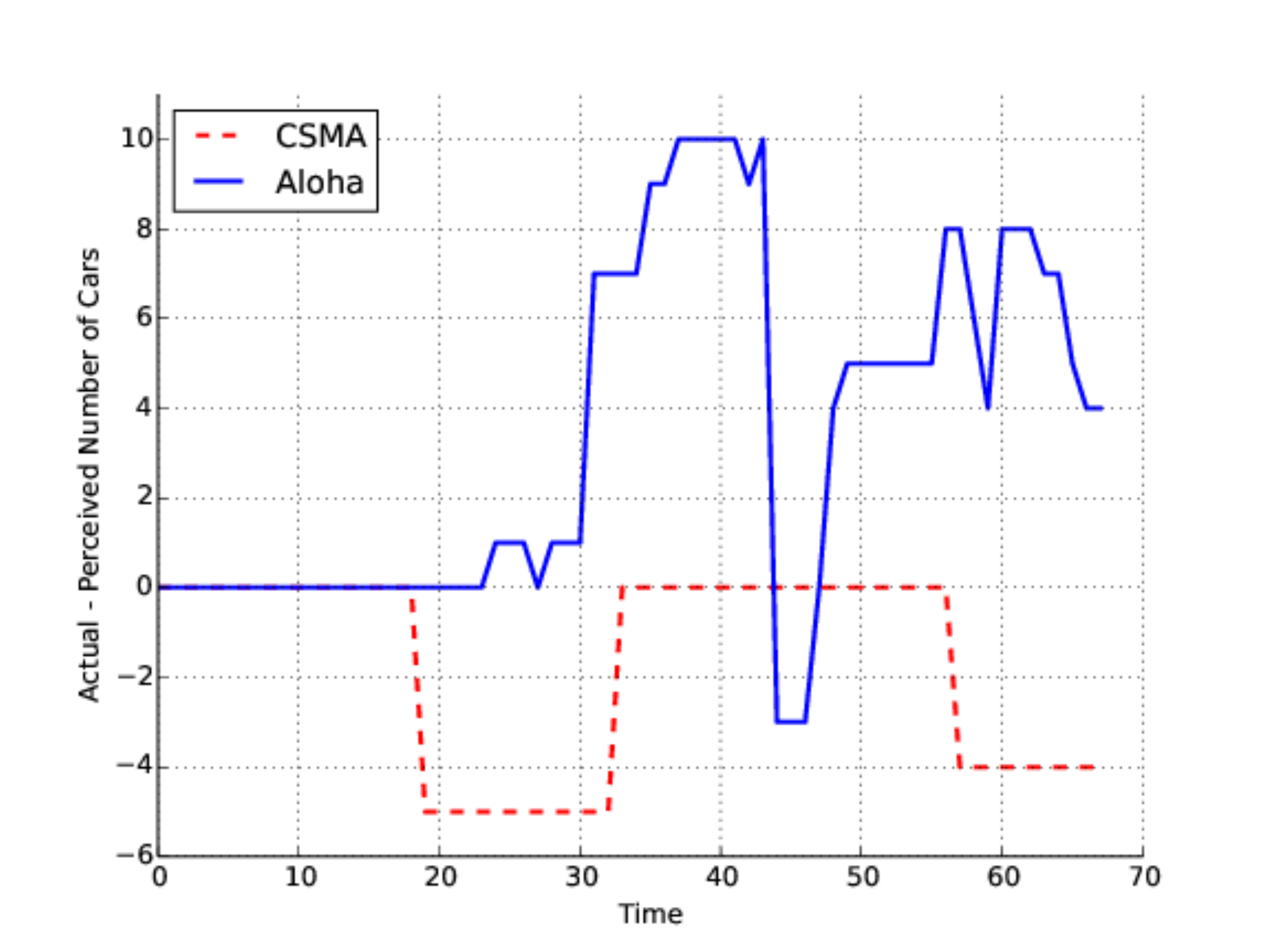}
  \caption{\small{Impact of MAC on Perception of Situation on a
    single-intersection scenario. Vehicles always travel in a straight
    line at constant speed, unless they need to stop due to traffic
    lights or other cars. At each iteration the probability of a new
    car arriving at one of the four edges of the grid and travelling in
    the corresponding direction is $50\%$.}}
  \label{fig:rvpdiff}
  \vspace{-0.5cm}
\end{figure}

Notably, the semantics of communications greatly impact the operation of the physical system. Fig. \ref{fig:rvpdiff} exemplifies this notion through a depiction of the difference between the actual number of cars waiting at a traffic light and the perceived number of cars known to the DM component of the system. As revealed by ABM, the minor difference of pre-sensing a channel (CSMA) or not (Aloha) causes either an over- or an under-estimation of the actual number of vehicles by the controlling element in the system. As such, the application of ABM techniques allows the development of an understanding of the various inter-relationships that direct the behavior of a complete telecommunication system.

\subsubsection{Functional Complexity}

As another example of work at the modelling layer, we have developed a metric to capture the amount of information shared between elements of a network (as a result of the organization of the network) in support of a network function. This analytical approach to quantify the complexity of a functional topology provides us with the means to capture the signaling complexity of functional operations within a network, such as handover or frequency assignment. That is, our complexity metric provides a new method of describing the functional operation of telecommunication networks.

Our complexity metric is built upon the concept of Shannon entropy ($H_r(x_n)$). We employ the Bernoulli random variable $x_n$ to model the potential of a node to interact with other nodes. The probability of interaction $p_r(x_n=1)$ is defined as the reachability of a node $n$ ($p_r(x_n=1) = i_r^n/j$, where $i_r^n$ is the number of nodes that can reach node $n$ and $j$ is the number of nodes for the given subgraph). The definition of reachability, in terms of the number of hops allowed between two nodes in the functional topology, enables the analysis of complexity on multiple scales ($r$). The one hop reachability represents the lowest possible scale ($r = 1$), where each node interacts only with its immediate neighbors. The increasing number of allowed hops between the nodes brings the nodes closer to each other in terms of interactions, and moves the focus from interactions among nodes to interactions among groups of nodes, i.e., analysis of higher scales. The total amount of information of the $k^{th}$ subgraph with $j$ nodes for scale $r$ is calculated as
	\begin{eqnarray}\label{eq:totalAmountOfInformationSubgraph}
	I_r(\Lambda_k^j) = \displaystyle\sum_{n \in \Lambda_k^j} H_r(x_n),
	\end{eqnarray}
where $\Lambda_k^j$ is the $k^{th}$ subgraph with $j$ nodes. The total amount of information represents the total uncertainty which is related to the actual roles of nodes that appear within a subgraph and different subgraph patterns. Our complexity metric, which is calculated with Eq. (\ref{eq:complexity}), quantifies the amount of order and structure in a system that is seemingly disordered.
	
	\begin{eqnarray}\label{eq:complexity}
	C_F = \dfrac{1}{R-1}\displaystyle\sum_{r =1}^{R-1}\sum_{j = 1+r}^{N} | \langle I_r(\Lambda^j) \rangle - \dfrac{r+1-j}{r+1-N}  I_r(\Lambda^N)|
    \label{CF}
	\end{eqnarray}
	
where $R$ is the maximum scale size, which is defined as the diameter of the functional topology, $N$ is the number of nodes in the functional topology, $\Lambda^N$ is the whole functional graph, and $\langle I_r(\Lambda^j) \rangle$ is the average amount of information for a given subgraph size $j$. We call the metric in Eq. (\ref{CF}) \emph{functional complexity}.
	
Our approach holistically gauges the functional organization of a network by first describing the interactions necessary to perform a given function topologically. Within this representation, we capture the network elements involved in performing some function and the interactions that support the operation of the function.

\begin{figure}
  \centering
	\includegraphics[width=8cm, height=5.5cm, trim={90 40 20 40}]{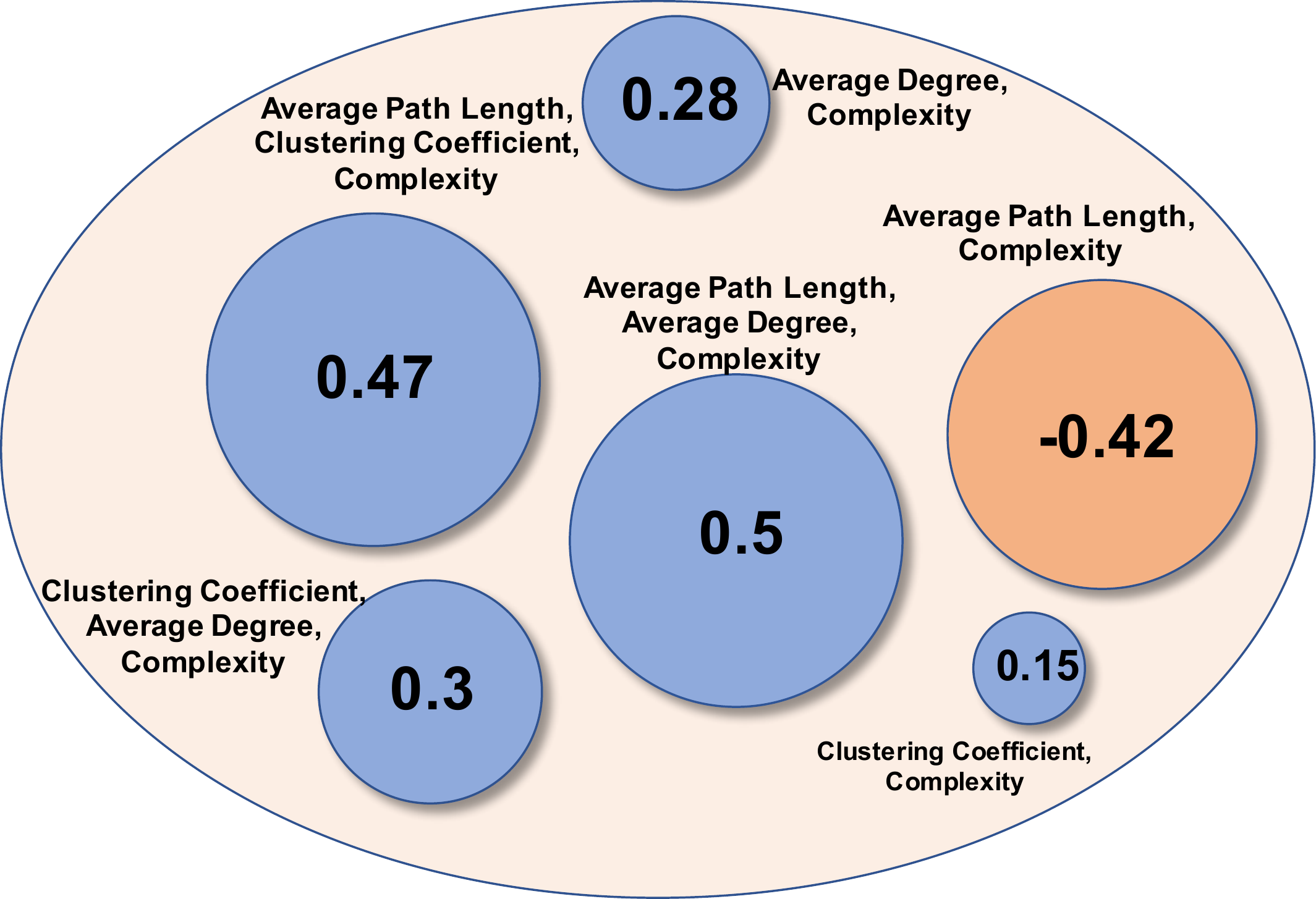}
  \vspace{0.5cm}
  \caption{\small{Correlation between the proposed complexity metric and the three most used measures of network topology (i.e., average path length, average degree distribution, clustering coefficient).}}
  \label{fig:correlation}
  \vspace{-0.5cm}
\end{figure}

Our quantification of networks in terms of their functional relationships provides a wholly new approach to understanding the operation of networks. As corroborated by Fig. \ref{fig:correlation}, more typical metrics for network topology do not capture the notions represented by our complexity metric (in fact, the correlation of our complexity metric with other traditional metrics is lower than 0.5 in all the cases we consider); this complexity metric thus provides an alternative method of describing network operation.

The above functional topology and complexity framework can be applied for instance to understand the underlying mechanisms that lead to certain network properties (i.e., scalability, energy efficiency) in Wireless Sensor Networks (WSN) as the result of different clustering algorithms \cite{Dzaferagic2017syscon}.

\subsection{Analysis Layer}

In the context of the analysis layer of our framework, we focus on a cellular network that self-organises from a frequency perspective to understand the collective behaviour of the network. We calculate the \emph{excess entropy}
\begin{eqnarray}\label{eq:excess_entropy}
E_C = \displaystyle\sum_{M = 1}^{\infty} (h(M) - h)
\end{eqnarray}	
to measure complexity, and the entropy
\begin{eqnarray}\label{eq:entropy_density}
h = \displaystyle\lim_{M \to \infty} h(M),
\end{eqnarray}
where $h(M)$ is the entropy of the target cell $X$ conditioned on $M$ surrounding cells. By measuring $E_C$ and $h$ we gain an understanding of the structure emerging in the lattice for a self-organising network.

Based on Eqs. (\ref{eq:excess_entropy}) and (\ref{eq:entropy_density}), in \cite{Macaluso2016} one shows that a self-organising cellular network can exhibit a complex behaviour, and that it can be robust against changes in the environment. In more detail, a self-organised and a centralised channel allocation are analyzed, with respect to their robustness to local changes in the environment. In order to compare the stability of the two types of channel allocation, $10^2$ instances of the self-organising frequency allocation algorithm are run using $10^2 \times 10^2$ lattices. Then, for each resulting channel allocation, all possible cells $n$ are considered, and for each cell all possible frequencies are in turn considered. Then the optimal minimum distance $c$ to an interference-free channel allocation is computed (we define the distance between two channel allocations as the number of changes that are necessary to move from one configuration to the other). We found that the locally perturbed channel allocation matrices resulting from self-organisation are more stable than those resulting from a centralized frequency planner.

What we know so far is that there is a relation between some complexity metrics and some telecom KPIs (i.e., between excess entropy and robustness to changes \cite{Macaluso2016}, and between functional complexity and the trade-off scalability-energy efficiency \cite{Dzaferagic2017syscon}). The complexity metrics we introduced have shed some new light on very relevant telecom KPIs/properties in the context of 5G networks, i.e., excess entropy can measure self-organization capabilities in the frequency allocation context and functional complexity can measure scalability in WSN. As widely acknowledged, \emph{self-organization and scalability} are very important properties of 5G systems (e.g., for IoT and dense small cell deployments). In the future we plan to improve and expand such understanding to all the most prominent 5G network technologies and KPIs.

\subsection{Tuning Layer}

ABM rules will choose the technological behaviour options that maximize the targeted communication network KPI, subject to constraints defined by the correlation between CSS metrics and other telecom KPIs. Local decisions will be based on only a few CSS metrics, and will lead to desired global behaviours/KPIs of the network. The local decisions are made according to ABM rules, by exploring and selecting the fittest behaviours (where by behaviour we mean some algorithm or policy acting on some radio resources).

\begin{figure}
	\centering
	\includegraphics[width=9cm]{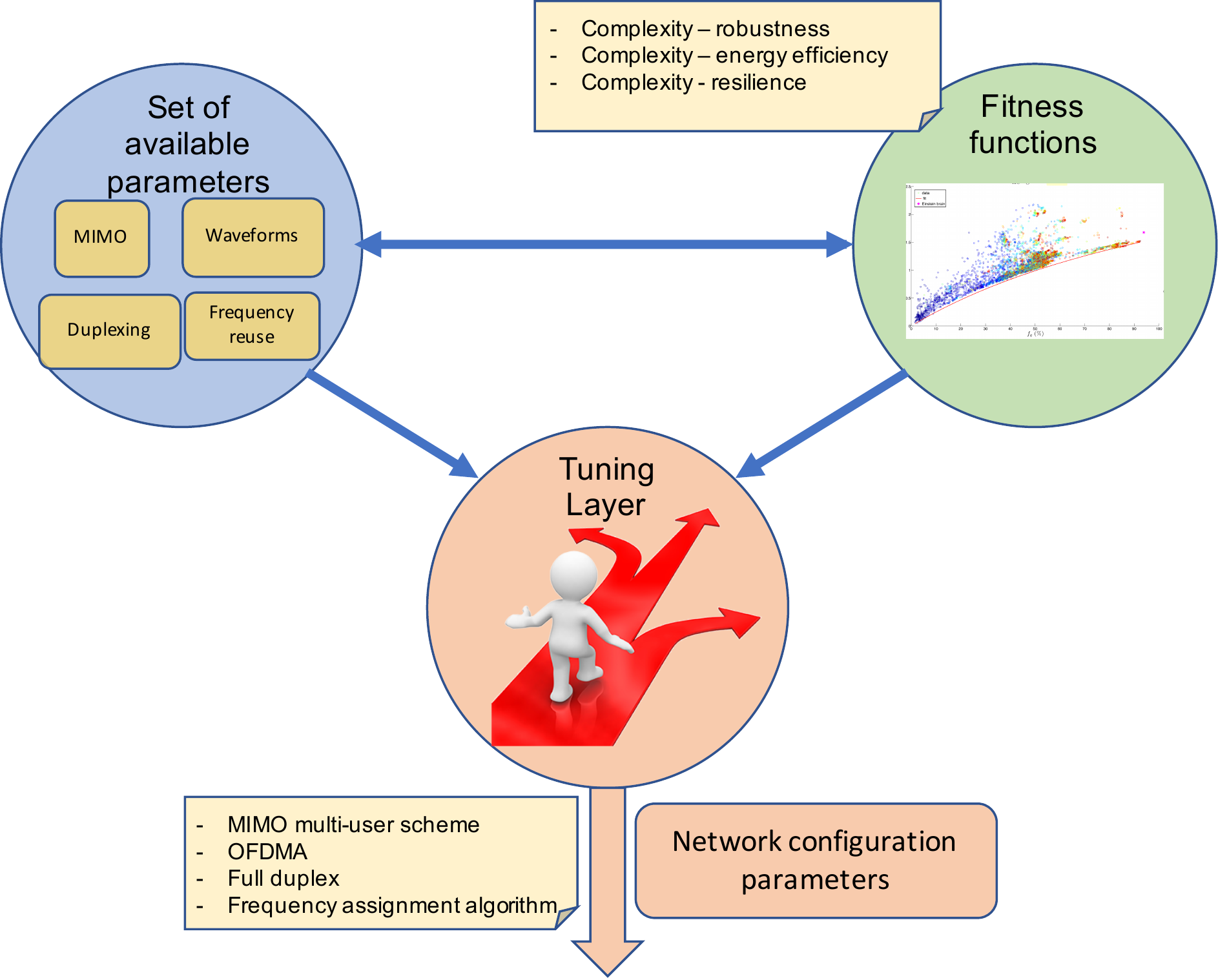}
	\caption{\small{The adaptation of network configuration parameters in the tuning layer. The set of available parameters represents a virtual pool of all the available network resources. The fitness functions depict a relationship between different network KPIs and complexity metrics which are calculated upon the set of available parameters.}}
	\label{fig:tuning}
\vspace{-0.5cm}
\end{figure}

Our goal, for different services (e.g., mobile broadband, M2M) is to choose behaviours that allow the network to achieve satisfactory KPIs, in terms of, e.g., delay, throughput, coverage, energy efficiency, out-of-band emission, etc. The question is whether we can keep achieving globally satisfactory KPIs just by changing ABM rules in a distributed fashion at different nodes. Such adaptation will act within a certain resource allocation domain (e.g., picking among different massive MIMO schemes) or between performance-equivalent allocations using resources from different domains (e.g., spectrum or infrastructure). The main ideas behind the tuning layer of our framework are exemplified in Fig. \ref{fig:tuning}.

Although our own work on the tuning layer is still in the initial phase, from a substantial amount of literature, we can gather evidence that different physical layer (PHY) and Radio Resource Management (RRM) techniques in the 5G domain should be chosen depending on environmental conditions and network requirements, i.e., we are potentially in a situation where our tuning layer is relevant and beneficial. We give a brief account of such evidence next.

In \cite{Bentosela2016}, it is shown that for a massive MIMO system, the sum-rate has a linear or sublinear behaviour with respect to the number of base station antennas, depending on the spatial richness of the environment; related work on adaptive precoding for distributed MIMO is explored in \cite{Ryu2015}. Several works investigate the coexistence of various waveforms in terms of cross-waveform leakage interference \cite{Xing2014, Bodinier2016} and possible implications for the waveform selection \cite{Sexton2016}. The fraction of cells that have full duplex base stations can be used as a design parameter, to target an optimal trade-off between area spectral efficiency and outage in a mixed full/half duplex cellular system \cite{Goyal2016, Cirik2015}. In \cite{Galiotto2017}, it is shown that increasing the frequency reuse can improve the throughput-coverage trade-off for ultra-dense small cell deployments, while a lower frequency reuse should be favoured if the target is maximizing throughput given a certain BS density.

In summary, we plan to use the above understanding of the benefit of adaptation at PHY and MAC layers in 5G networks, and extend it as needed in terms of technology components, KPIs and adaptation criteria, to inform our framework, and show its immediate benefit in understanding, operating and designing 5G systems.

%% file: OpenChallenges_5.tex
\section{Open Challenges}
\label{future-work}

Several more \emph{5G component technologies}, in addition to those considered in this paper, can enrich the set of possible choices used to model, analyse and tune the network, including (massive) co-located or distributed multiple antenna arrays; different waveforms and multiple access schemes; different duplexing schemes; novel spectrum sharing schemes such as License Assisted Access (LAA); and different frequency reuse schemes including probabilistic ones for ultra-dense networks.

What we know so far is that there is a relation between some complexity metrics and some telecom KPIs (i.e., between excess entropy and robustness to changes, and between functional complexity and the trade-off scalability-energy efficiency). In the future the aim is to improve and expand such understanding to all the most prominent technologies and KPIs for 5G networks. In particular it is still an open question how to achieve the desired network tuning properties within a large optimization space encompassing many different network resources, KPI objectives and constraints, many different heterogeneous co-existing networks and a very large number of nodes and decision points. We conjecture ABM can help us achieve such ambitious goal, as a key tool to engineer desired emergent properties in such future challenging networks.

As the network graph representations discussed in the proposed framework might dynamically change according to the different radio resource domains and related techniques used, one open area of investigation is to study how the complexity metrics can be calculated and how they evolve over time for such dynamic multi-dimensional resource allocation; and then use such metrics to analyse and tune the network behaviour taking into account robustness, resilience, network utilization, and other \emph{time-dependent network characteristics}. 

%% file: Conclusion_5.tex
\section{Conclusion}
\label{conclusion}

Current complex systems science literature focusing on communication systems draws on network science, studying applications and traffic modelling, but lacks considerations of architecture, infrastructure, and technology. We instead apply complex systems science to wireless networks from the \emph{functional perspective}, drawing on concepts from information (i.e., different metrics of complexity) and computational (i.e., agent based modeling) theory, adapted from complex system science.


Since complex systems science metrics are currently absent from the quantities considered when operating and designing communication networks, by introducing our proposed framework we initiate a completely new way to model, analyse and engineer networks, founding a new theory and practice of telecommunications not previously anticipated. As a simple example, our work on exploring frequency planning from a complex systems perspective leads us to conclude that future networks shall eschew any current frequency planning approaches and instead determine frequency of operation on the fly, with enormous implications for design, roll-out and operation of networks. We believe such distributed decision making paradigm is likely going to be the way forward for many of the future 5G and IoT resource allocation problems. In particular, we have reasons to believe that complex systems science provides the key to unlock the full potential of self-organization in telecom systems.